\documentclass[a4paper,fleqn,usenatbib]{mnras}
\usepackage{lscape,graphicx}
\usepackage{rotating}
\usepackage{color}

\usepackage{color}

\def\apj{ApJ}
\def\apjs{ApJS}
\def\mnras{MNRAS}

\def\apjl{ApJ}
\def\aap{A\&A}
\def\aj{AJ}

\def\araa{ARA\&A}

\def\gca{Geochim. Cosmochim. Acta}
\def\pasp{PASP}

\def\asec{\ifmmode ^{\prime\prime}\else$^{\prime\prime}$\fi}

\def\it{\sl}
\def\degs{\ifmmode ^{\circ}\else$^{\circ}$\fi}
\def\amin{\ifmmode ^{\prime}\else$^{\prime}$\fi}
\def\asec{\ifmmode ^{\prime\prime}\else$^{\prime\prime}$\fi}
\def\fdg{\hbox{$.\!\!^\circ$}}          
\def\farcs{\hbox{$.\!\!^{\prime\prime}$}}  

\def\degs{\ifmmode ^{\circ}\else$^{\circ}$\fi}
\def\amin{\ifmmode ^{\prime}\else$^{\prime}$\fi}
\def\farcm{\hbox{$.\mkern-4mu^\prime$}}
\def\eqalign#1{\null\,\vcenter{\openup1\jot \m@th
   \ialign{\strut\hfil$\displaystyle{##}$&$\displaystyle{{}##}$\hfil
   \crcr#1\crcr}}\,}

\title[\textit{XMM-Newton} and \textit{Chandra} observations of DA~495]{Constraining the parameters of the pulsar wind nebula DA~495 and its pulsar with \textit{Chandra} and \textit{XMM-Newton}  
}
\author[A.~Karpova, D.~Zyuzin, A.~Danilenko, Yu.~Shibanov]{
A.~Karpova$^{1,2}$\thanks{E-mail: annakarpova1989@gmail.com}, 
D.~Zyuzin$^{1}$, 
A.~Danilenko$^{1}$, 
Yu.~Shibanov$^{1,2}$\\
$^{1}$Ioffe Institute, Politekhnicheskaya 26, St. Petersburg, 194021, Russia\\
$^{2}$Peter the Great St.Petersburg Polytechnic University, Politekhnicheskaya 29, St. Petersburg, 195251, Russia}

\date{Accepted XXX. Received YYY; in original form ZZZ}
\pubyear{2015}

\begin{document}

\label{firstpage}
\pagerange{\pageref{firstpage}--\pageref{lastpage}} 
\maketitle

\begin{abstract}   
We present  spectral and timing analyses of the X-ray emission 
from the pulsar wind nebula DA~495 and its central object, J1952.2+2925, 
suggested to be the pulsar, using archival \textit{Chandra} and \textit{XMM-Newton} data.
J1952.2+2925 has a pure thermal spectrum which is equally well fitted
either by the blackbody model with a temperature of $\approx 215$~eV and 
an emitting area radius of $\approx 0.6$~km 
or by magnetized neutron star atmosphere models with temperatures of 80--90~eV.     
In the latter case the thermal emission can come from the entire neutron star surface which temperature 
is consistent with standard neutron star  cooling scenarios. We place also an upper limit 
on the J1952.2+2925 nonthermal  flux.   
The derived spectral parameters  are generally compatible  with published ones based only 
on the \textit{Chandra} data, but they are much more accurate due to the inclusion of  \textit{XMM-Newton} data.
No pulsations were found and we placed  an upper limit for 
the J1952.2+2925 pulsed emission fraction of  40~per~cent. 
Utilizing the interstellar absorption--distance relation, we estimated the distance to DA~495, 
which can be as large as 5~kpc if J1952.2+2925 emission is described by the atmosphere models. 
We compiled possible multi-wavelength spectra of the nebula including radio data; they depend on 
the spectral model of the central object.
Comparing the results with other pulsar plus wind nebula systems we set reasonable constraints 
on the J1952.2+2925 spin-down luminosity and age.   
We suggest that the \textit{Fermi} source 3FGL J1951.6+2926 is the likely $\gamma$-ray counterpart 
of J1952.2+2925. 
\end{abstract}


\section{Introduction}

More than 50 pulsar wind nebulae (PWNe) are currently known \citep[e.g.,][]{kargaltsev2008}. 
The Crab nebula is a classical example of  young PWNe which is the best studied  from the radio to the TeV range.
However, other PWNe, not yet studied in such details, show a variety of individual properties 
different from the Crab, which are determined by their  pulsar  parameters, evolution stages 
and surroundings, including the interstellar medium and the supernova ejecta.    
Studying such PWNe is crucial for understanding the properties of their central `engines', 
the interaction of the relativistic pulsar wind with surroundings and particle acceleration 
mechanisms \citep{gaensler2006}.
Multiwavelength observations of  the PWNe at late evolutionary stages and different environment 
are still rare but they are particularly important  for comparison  with the Crab and various theoretical models 
of  evolution of these objects \citep[and references therein]{gelfand2009, tanaka2010, 2011chevalier}.     

\begin{figure}
\begin{minipage}[h]{1.\linewidth}
\center{\includegraphics[width=0.80\linewidth, trim={0 0 2cm 2cm}, clip]{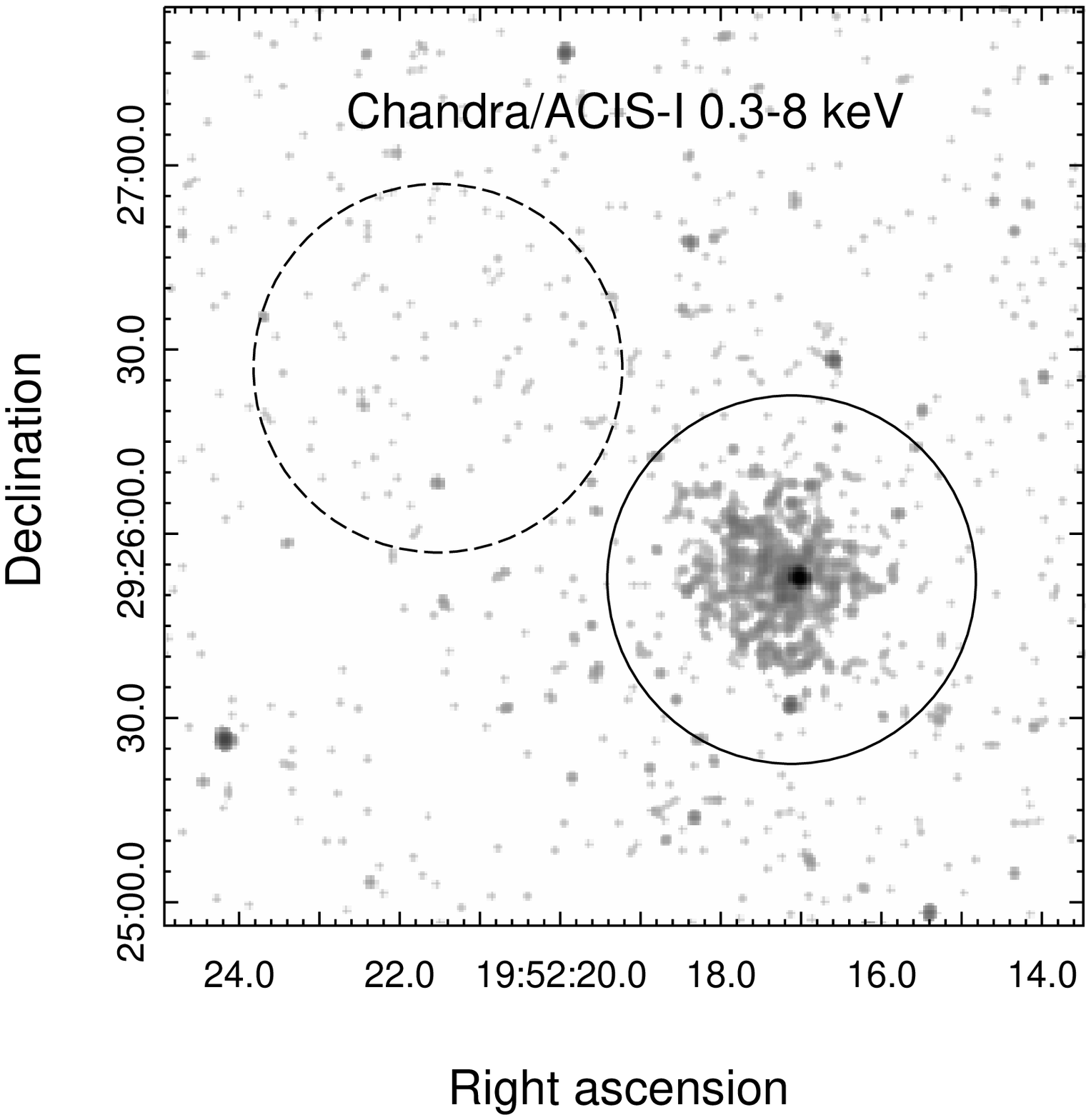}}
\end{minipage}\\
\begin{minipage}[h]{1.\linewidth}
\center{\includegraphics[width=0.80\linewidth, trim={0 0 2cm 2cm}, clip]{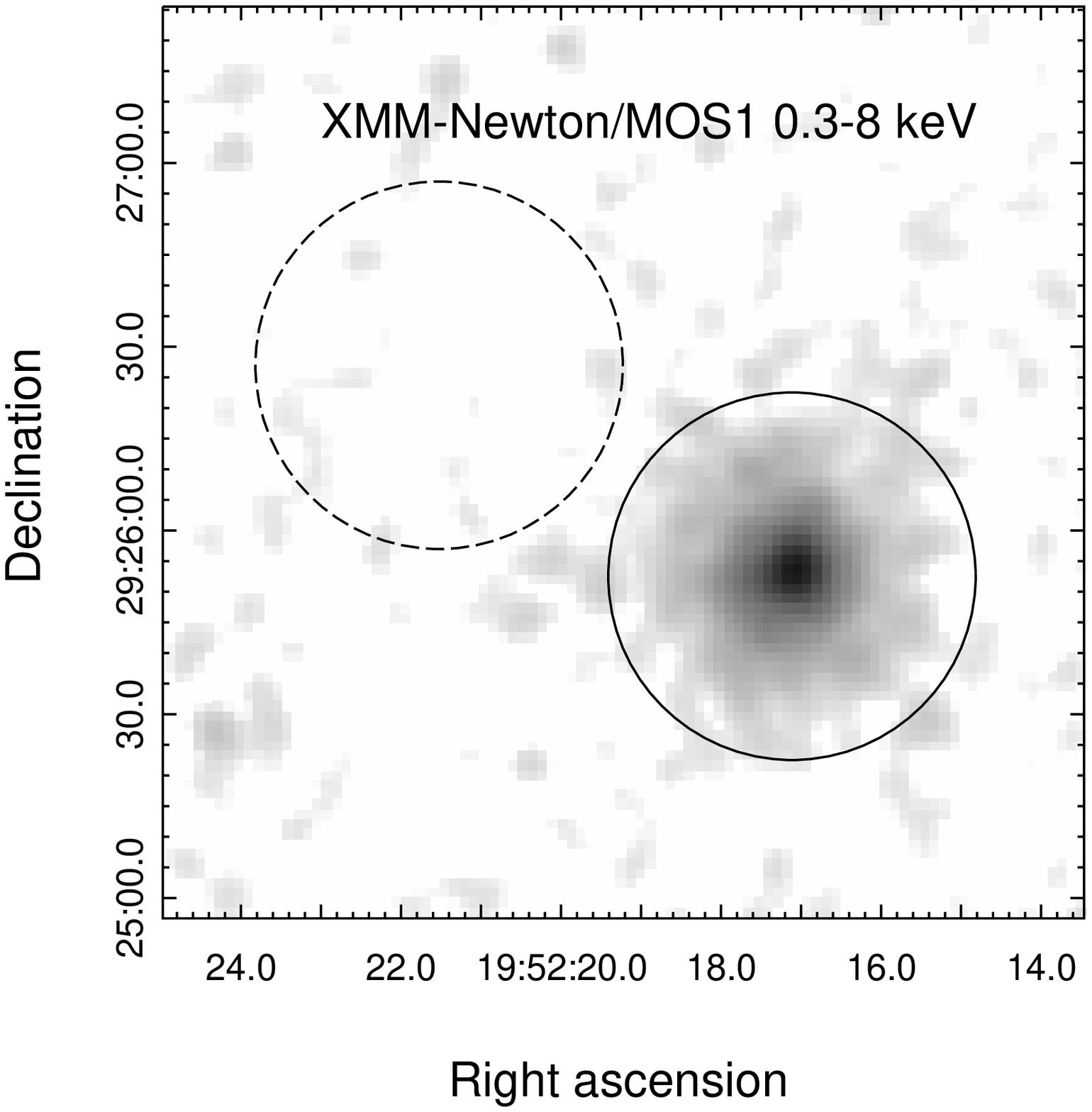}}
\end{minipage}\\
\begin{minipage}[h]{1.\linewidth}
\center{\includegraphics[width=0.77\linewidth, trim={0.2cm 0 2cm 2cm}, clip]{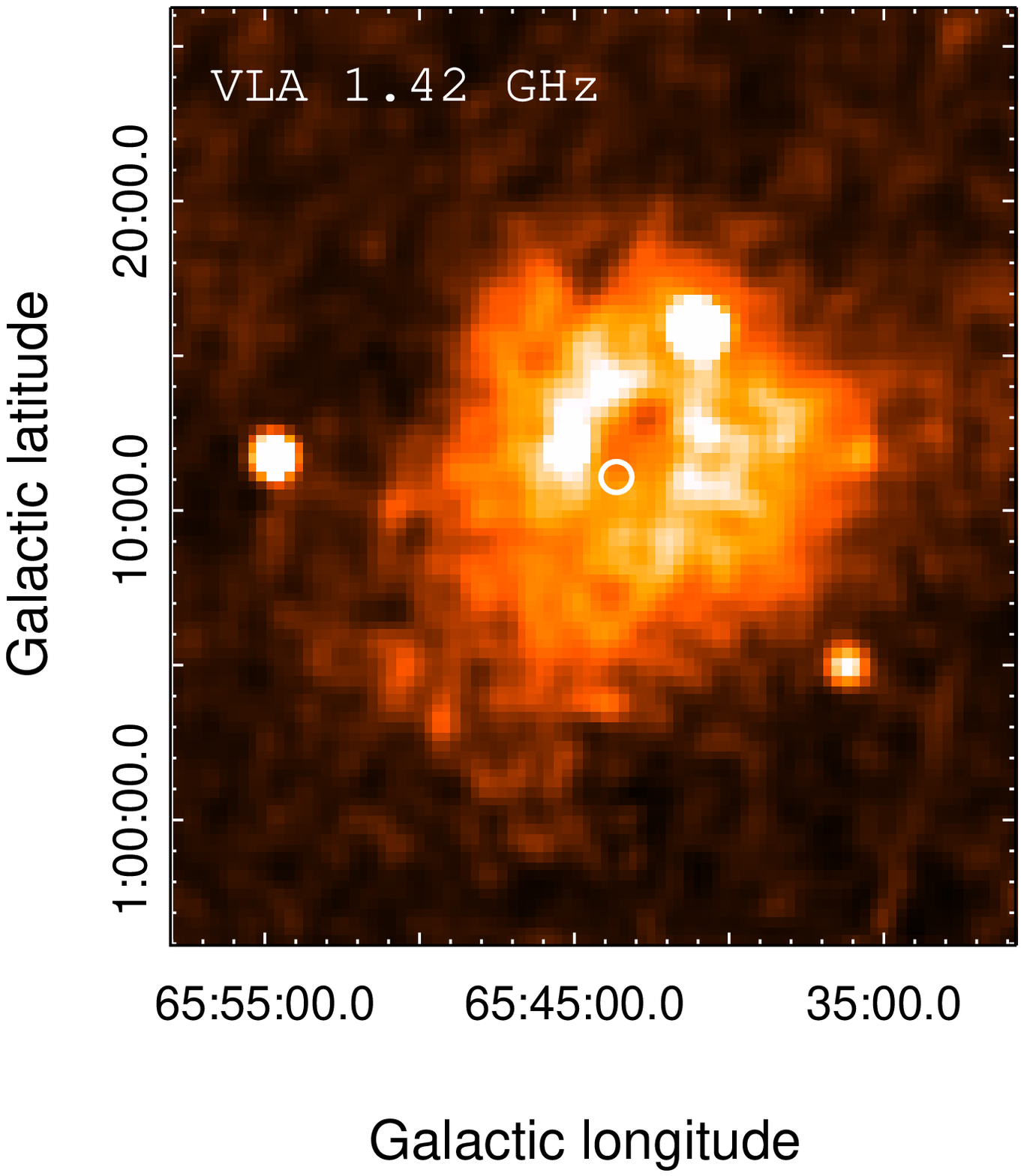}}
\end{minipage}
\caption{2\farcm5$\times$2\farcm5~X-ray \textit{Chandra} ({\sl top}), \textit{XMM-Newton} ({\sl middle}) 
  and 30\amin$\times$27\amin~radio VLA ({\sl bottom}) images of DA~495 in 0.3--8 keV range and at 1.42 GHz, respectively. 
  The \textit{Chandra} image was smoothed with a 3 pixel 
  Gaussian kernel (logarithmic brightness scale is used). 
  J1952 is clearly seen in the centre of the PWN in X-rays. 
  The \textit{XMM-Newton} image was binned to a pixel size of 1\farcs6 and smoothed with a 3 pixel 
  Gaussian kernel (square root brightness scale is used). Solid and dashed circles
  depict the extraction region for the PWN+J1952 and background, respectively. 
  The white circle in the radio image shows the position and extent of DA~495 in X-rays.}
\label{fig:img}
\end{figure}

DA~495 (G65.7$+$1.2) appears to be an example of an evolved PWN. 
It was discovered in the radio \citep{galt1968}.
Detailed radio observations showed that DA~495 has an uncommon annular morphology with a central 
flux deficit within $\sim$2~arcmin from the centre, dubbed the radio `hole' \citep{landecker1983}. 
The full extent of DA~495 is about 25 arcmin and the radio emissivity decreases gradually to its outer edge 
without any evidence of the supernova remnant (SNR) shell existence 
(see Fig.~\ref{fig:img}, bottom panel).  
A possible association with the open cluster NGC 6834 implied a distance $D$ to the object 
of $\approx2.1$~kpc \citep{landecker1983}.
H{\sc\,i} absorption measurements and  a flat rotation model for the Milky Way provided $D$ within a range of
1--1.5~kpc, depending on  the adopted Galactocentric radius for the Sun \citep{kothes2004,kothes2008}.  
\citet{kothes2008} found a spectral break at 1.3~GHz, the lowest for the known PWNe,
with spectral indices $\alpha_{\nu}=0.45\pm0.20$ and $\alpha_{\nu}=0.87\pm0.10$ below and above the break,
respectively\footnote{Assuming the spectral flux $F_{\nu}\propto \nu^{-\alpha_{\nu}}$, 
where $\nu$ is the frequency.}. 
They also suggested that DA~495 is about 20 kyr old. 
No detection of an associated pulsar in the radio and $\gamma$-rays was reported.  

In X-rays, \textit{ROSAT} and \textit{ASCA} observations of DA~495 revealed a faint compact source, 
1WGA J1952.2$+$2925 (hereafter J1952), which is apparently surrounded by a diffuse 
non-thermal emission and is projected on the edge of the radio hole \citep{arzoumanian2004} (see Fig.~\ref{fig:img}). 
It was proposed to be a magnetospherically active neutron star (NS)  powering the PWN,  
although only an upper limit on the pulsed emission fraction of 50~per~cent for periods $\ga$~30 ms 
was derived from the analysis of the \textit{ASCA} data.  
The compact source J1952 and its diffuse emission were later firmly confirmed by 
\textit{Chandra} high spatial resolution observations \citep{arzoumanian2008}. 
It was established that J1952 is the point source     
located in the centre of the X-ray nebula with an extent of $\sim$~40\arcsec. 
The nebula does not show any Crab-like torus$+$jet structure, however its spectrum is described 
by a power law with a photon index $\Gamma=1.6\pm0.3$ typical for PWNe. The latter allowed to state    
that it is the X-ray counterpart of the DA~495 PWN.           
J1952, presumably the pulsar, has a pure thermal spectrum. It can be described either 
by the blackbody model with a temperature $T\approx2.5$ MK and an emitting area radius  
$R\approx 0.3$~km or by the neutron star atmosphere (NSA) model \citep*{nsa1996} for a NS with 
the effective temperature of $\approx1$ MK and the radius of 10 km. 
The interstellar absorption column density for the former and latter cases was 
$\approx2.3 \times10^{21}$ and $6.0 \times10^{21}$~cm$^{-2}$, respectively, which is a factor 
of 1.3--3.5 lower than the entire Galactic absorption in this direction. 
Because of small count statistics, the spectral parameters were poorly constrained.  
In addition, two key parameters, the distance and the pulsar spin-down energy 
loss $\dot{E}$, also remained uncertain, which did not allow to establish firmly 
the DA~495 PWN evolution stage. Finally, a high energy source 3FGL J1951.6+2926 recently detected 
with \textit{Fermi}/LAT was proposed as a possible $\gamma$-ray counterpart \citep{3Fermi} of the nebula.  

Here we report a simultaneous analysis of the 
\textit{Chandra}\footnote{PI Arzoumanian, \textit{Chandra}/ACIS-I, ObsID 3900}  
and  unpublished  \textit{XMM-Newton}\footnote{PI Arzoumanian, \textit{XMM-Newton}/EPIC, ObsID 0406960101} 
X-ray archival data on  DA~495. 
We also include in our analysis  the extinction--distance relation towards J1952.  
This allows us to improve considerably the count statistics, to get an independent  
DA~495 distance estimate and to set more stringent constraints on the PWN and pulsar parameters.
We also use high temporal resolution \textit{XMM-Newton}/EPIC-pn data to search for periodic 
pulsations from J1952 and derive a more stringent upper limit on its pulsed emission fraction. 
The details of observations are described in Sect.~\ref{sec:data}. 
The timing analysis is presented in Sect.~\ref{sec:timing}. 
The extinction--distance relation and spectral analysis are described 
in Sect.~\ref{sec:relation} and \ref{sec:analysis}. 
In Sect.~\ref{sec:discus} we discuss our results, compare them with the DA~495 radio and $\gamma$ data 
and the data for other pulsar+PWN systems. 
A summary is given in Sect.~\ref{sec:sum}.

\begin{figure}
\centerline{
\includegraphics[scale=0.41]{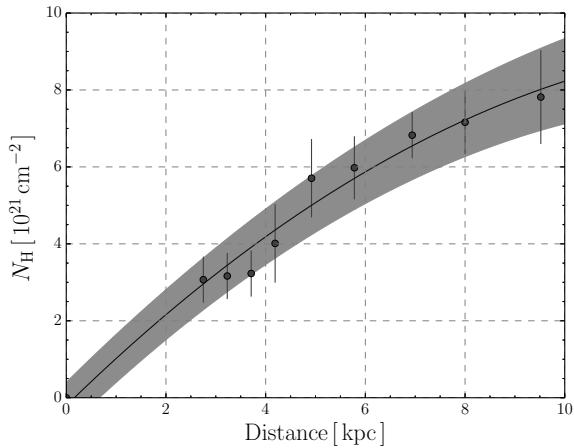}}
  \caption{Data points show empirical $N_{\rm H}$--distance relation for the DA~495 direction
  derived using \citet{marshall2006} extinction map. 
  The error bars represent 1$\sigma$ uncertainties.
  The solid line and gray filled region are smoothing spline
  approximations to the data points and their uncertainties, respectively.}
\label{fig:Nh-d}
\end{figure}

\section{The X-ray data}
\label{sec:data}

The \textit{XMM-Newton} observations of DA~495 were 
carried on 2007 April 21 with total exposure of about 50 ks. 
The EPIC-MOS cameras were operated in the Full Frame Mode with the medium filter setting, 
and the EPIC-pn camera was operated in the Small Window Mode with the thin filter. 
The {\sc xmm-sas}~v.13.5.0 software was used to process the data. 
We selected single and double pixel events (PATTERN~$\leq$~4) for the EPIC-pn and single 
to quadruple-pixel events (PATTERN~$\leq$~12) for the EPIC-MOS data. 
We removed periods of background flares using 10--12 keV and 12--14 keV light curves for 
the MOS and pn data, respectively. 

We also used the observation of DA~495 made with \textit{Chandra}/ACIS-I on 2002 December 9. 
The data mode was VFAINT, the exposure mode was TE and the total exposure was 24.8 ks. 
The {\sc ciao}~v.4.6 {\sc chandra\_repro} tool with {\sc caldb}~v.4.6.1 was used to reprocess the data. 
We found no background flares in this observation.
The resulting effective exposures are presented in Table~\ref{t:counts}.

\section{Timing analysis}
\label{sec:timing}

The EPIC-pn data, which have the time resolution of 5.6 ms, were used to search for pulsations from J1952. 
Event times were corrected to the solar system barycentre using the {\sc sas} task {\sc barycen} and J1952  
\textit{Chandra} coordinates obtained with the {\sc ciao} tool {\sc wavdetect} 
(RA = 298\fdg0710, Dec = 29\fdg4313). 
We used a circular region of 20~arcsec around J1952 to extract its light curve. 
The {\sc sas} task {\sc epiclccorr} was used for background subtraction and 
correction for vignetting, bad pixels and quantum efficiency. 
The resulting light curve contained about 1060 source counts in the 0.5--1.4 keV energy band. 
We chose this energy range to minimize the contribution of the PWN emission and
since J1952 dominates at these energies (see Sect.~\ref{sec:analysis}). 
After that we performed a Fast Fourier Transform (FFT) analysis using standard {\sc xronos 
v.5.22}\footnote{http://heasarc.gsfc.nasa.gov/docs/xanadu/xronos/xronos.html} task {\sc powspec}.
No pulsations were found. 
We derived an upper limit for the pulsed fraction (PF) of 40~per~cent (at the 99~per~cent confidence level) 
for periods $\ga12$~ms following the procedure described by \citet{vaughan1994}. 

\section{Distance and interstellar extinction}
\label{sec:relation}

It is possible to estimate the distance to DA~495 using the absorption column 
density $N_{\rm H}$--distance relation along its line of sight. 
To do this we use the interstellar extinction distribution obtained by
\citet{marshall2006} for the direction towards DA~495.
They used the 2MASS survey along with the Besan\c{c}on model of population synthesis \citep{robin2003} 
to calculate the $K_s$ extinction dependence on distance along different lines of sight. 
We obtained this dependence from the VizieR  
Service\footnote{http://vizier.u-strasbg.fr/viz-bin/VizieR-3?-source=J/A\%2bA/453/635/}
for the DA~495 direction and then transformed the extinctions to the column densities $N_{\rm H}$ 
utilizing a relation $A_{\rm K_s}/A_{\rm V}\approx0.1$ \citep[see e.g.][]{rieke1985} and a standard 
empirical $N_{\rm H}$--$A_{\rm V}$~relation \citep{predehl1995AsAp}.  
The resulting $N_{\rm H}$--distance dependence is shown in Fig.~\ref{fig:Nh-d}.  
The $N_{\rm H}$--distance relation and its uncertainties were approximated by smoothing splines.


\section{Analysis of the X-ray spectra}
\label{sec:analysis}

The \textit{Chandra} and \textit{XMM-Newton} images of DA~495 for 
the 0.3--8 keV energy range are shown in the top and middle panels of Fig.~\ref{fig:img}, respectively. 
In the \textit{Chandra} image the extended PWN is seen around the point source J1952. 
Due to a moderate angular resolution of \textit{XMM-Newton} the PWN+J1952 system appears to be 
slightly more extended\footnote{We have searched for a fainter PWN emission at larger spacial scales 
in the \textit{XMM-Newton} images but not found it.} and the point source is blended with the PWN.  
Therefore, to perform spectral analysis of J1952 and the PWN we used their individual spectra extracted 
from the \textit{Chandra} data, while from the \textit{XMM-Newton} data we extracted 
the spectrum corresponding to the combined PWN+J1952 emission. 
For the latter we chose 30~arcsec aperture encircling 
the whole PWN emission (see Fig.~\ref{fig:img}) using {\sc evselect} tool. 
We then used {\sc sas} tasks {\sc rmfgen} and {\sc arfgen} to generate the redistribution matrix and  
ancillary response files for the MOS1, MOS2 and pn detectors.  
For the \textit{Chandra} data set we extracted the J1952 spectrum from a circular region of 1.5-pixel~radius 
centred at this source and the PWN spectrum from the same circular region as for \textit{XMM-Newton} 
excluding 1~arcsec radius region around J1952 using {\sc specextract} tool.
Background was taken from the region free from any sources (see Fig.~\ref{fig:img}).
All spectra were grouped to ensure $\geq 5$ counts per energy bin. 
The resulting number of counts in the apertures and background-subtracted
source count rates are presented in Table~\ref{t:counts}.

\begin{landscape}

\begin{table}
\caption{Resulting {\textit{Chandra}} and {\textit{XMM-Newton}} exposure times, number of counts and 
background-subtracted count rates in 0.3--10 keV energy band for different instruments and source apertures 
for the J1952, PWN and J1952+PWN after data reprocessing, filtering and correction for background flares.}
\label{t:counts}
\begin{center}
\begin{tabular}{cccccc}
\hline
 Instrument   & \multicolumn{2}{c}{\textit{Chandra}}     & \multicolumn{3}{c}{\textit{XMM-Newton}}\\
 Detector   & \multicolumn{2}{c} {ACIS-I}     & \multicolumn{3}{c}{EPIC/}  \\ 
       & \multicolumn{2}{c} {}          & MOS1 &MOS2 &pn \\
 \hline
 Source & J1952 & PWN                                        & \multicolumn{3}{c}{J1952+PWN} \\ \hline
  exposure (ks) &24.8 &24.8                          &46.5 &47.5   &33.7 \\
  counts within the aperture (see text) &145 &577                            &1300 &1282 &4194 \\
  background-subtracted count rate (cts ks$^{-1}$) & 5.8$\pm$0.5 &19.6$\pm$1.0 &22.9$\pm$0.8  &22.2$\pm$0.8 &74.4$\pm$2.2 \\
  \hline
\end{tabular}
\end{center}
\end{table}

\begin{table}
\caption{Best-fit spectral parameters. 
All errors correspond to 90~per~cent credible intervals. 
$R$ and $T$ are given as seen by a distant observer. 
The PWN luminosity $L_{\rm pwn}$ is in the 0.5--8 keV range. $K_{\rm pwn}$ is the normalisation constant 
for the PL model of the PWN. 
$F_{\rm J1952}$ is the point source bolometric flux. $L^{\rm pl}_{\rm J1952}$ is a 3$\sigma$
upper limit on the non-thermal component of the point source luminosity in 0.5--8~keV band.}
\label{t:best-fit}
\begin{center}
\begin{tabular}{lllllllllll}
\hline
\multicolumn{10}{c}{} \\
Model &$N_{\rm H}$,        &$D$, &$R$, &$T$, &$\Gamma_{\rm pwn}$ &$K_{\rm pwn}$,    &Log$F_{\rm J1952}$, 
           &Log$L^{\rm pl}_{\rm J1952}$, &Log$L_{\rm pwn}$, &$\chi^2$/dof\\
& 10$^{21}$~cm$^{-2}$ &kpc  &km   &eV   & &10$^{-5}$ ph  & erg s$^{-1}$ cm$^{-2}$  
           &erg s$^{-1}$       &erg s$^{-1}$  &\\
    &       &     &     &     &               & keV$^{-1}$ cm$^{-2}$ & & & & \\    
\hline
\multicolumn{10}{c}{ }  \\
1. BB & $2.6^{+0.5}_{-0.4}$ & $2.4^{+1.3}_{-1.1}$ & $0.6^{+0.6}_{-0.3}$ & $215^{+23}_{-23}$ 
    & $1.71^{+0.12}_{-0.12}$  &$4.4^{+0.7}_{-0.6}$  &$-12.94^{+0.14}_{-0.11}$ & $<$ 31.51 
    & $32.22^{+0.38}_{-0.49}$ & 455/461 \\
\hline
\multicolumn{10}{c}{ }  \\
2. NSMAX & $3.5^{+0.7}_{-0.6}$ & $3.3^{+1.7}_{-1.3}$ & $10^{+21}_{-7}$      & $76^{+19}_{-16}$ 
    & $1.83^{+0.14}_{-0.13}$ & $5.1^{+0.9}_{-0.7}$ & $-12.47^{+0.35}_{-0.26}$ & $<$ 31.66
    & $32.53^{+0.37}_{-0.45}$ & 455/461 \\
$B=10^{12}$~G & & & & & & & & & & \\  
\hline
\multicolumn{10}{c}{ }  \\
3. NSMAX &$3.4^{+0.7}_{-0.6}$ & $3.2^{+1.5}_{-1.2}$ & $6^{+9}_{-3.5}$      & $91^{+17}_{-15}$ 
    & $1.82^{+0.13}_{-0.13}$ & $5.1^{+0.8}_{-0.6}$ & $-12.57^{+0.27}_{-0.20}$ & $<$ 31.72
    & $32.50^{+0.35}_{-0.42}$ & 454/461 \\
$B=10^{13}$~G & & & & & & & & & & \\ 
 \hline
\end{tabular}
\end{center}
\end{table}
\end{landscape}

A power law (PL) was used to model the PWN emission. 
For the J1952 spectrum, we tried the neutron star atmosphere models NSA \citep{nsa1996} and
NSMAX \citep*{ho2008ApJS} with a gravitational redshift $z_g+1=1.21$ and magnetic 
fields $B=10^{12}$~G and $B=10^{13}$~G and the blackbody (BB) model.
For the interstellar absorption, we used the {\sc xspec} photoelectric absorption {\sc phabs} model with 
default cross-sections {\sc bcmc} \citep{balucinska1992} and abundances {\sc angr} \citep{anders1989}.
We fitted \textit{XMM-Newton} and \textit{Chandra} spectra of 
the J1952+PWN, J1952 and PWN simultaneously in 0.3--10 keV energy range. 

\begin{figure*}
\begin{minipage}[h]{0.494\linewidth}
\center{\includegraphics[width=1.05\linewidth]{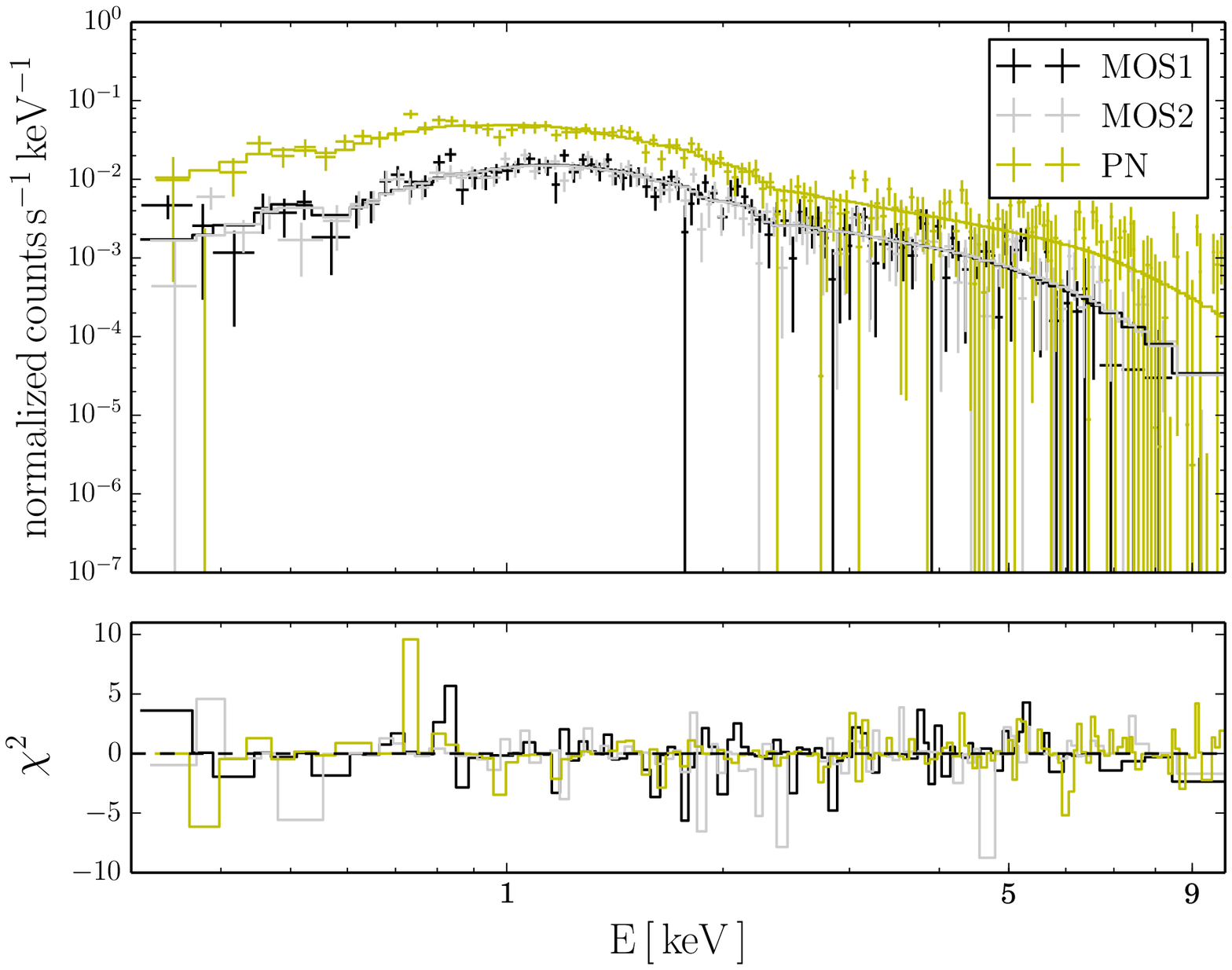}}
\end{minipage}
\hfill
\begin{minipage}[h]{0.494\linewidth}
\center{\includegraphics[width=1.05\linewidth]{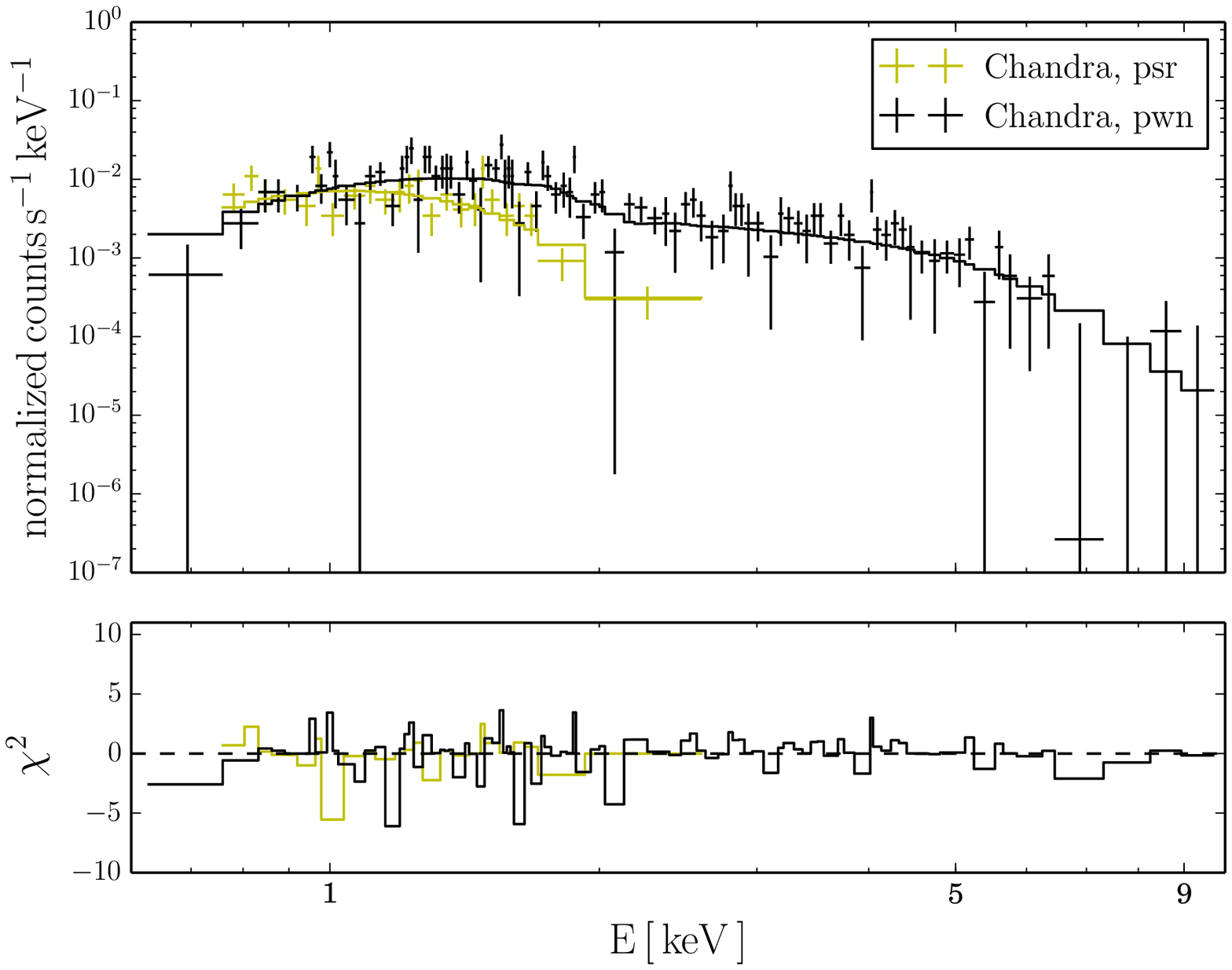}}
\end{minipage}
\caption{Observed X-ray spectrum of the DA~495 pulsar+PWN system obtained with {\it XMM-Newton} ({\sl top-left}) and {\it Chandra} ({\sl top-right}). The data for different instruments/sources are shown by different crosses, as indicated in the inserts. Lines show the best-fit model.
{\sl Bottom:} Fit residuals.}
\label{fig:spec}
\end{figure*}

We fit the data using the Bayesian inference with the $N_{\rm H}$--distance relation as a prior. 
We obtained samples from the posterior distribution by the Markov chain Monte-Carlo ({\sc mcmc}) sampler 
developed by \citet{goodman2010CAMCS} and implemented as a python package {\sc emcee} by \citet{foreman2013}. 
We obtained 100000 samples which is enough compared with autocorrelation time (50--80). 
The procedure is similar to that utilised by \citet{kirichenko2015}.
Posterior median values of the spectral parameters with 90~per~cent credible intervals 
are presented in Table \ref{t:best-fit}, where the models differ   
only by the thermal components applied and are named according to that. 
Since the best-fit parameters do not depend strongly on the specific type of atmospheric models, 
i.e., NSA or NSMAX, we present the results only for the BB and NSMAX models. 
The $\chi^2$ values per degree of freedom (dof) show that all models describe the data equally well. 
For both models our results are generally consistent with results of \citet{arzoumanian2008},  
however they are  several times less uncertain (cf. their Table 1) due to the better count statistics. 
The example of the fit for the model~2 from Table \ref{t:best-fit} is shown in Fig.~\ref{fig:spec}. 

It is important, that using the Bayesian approach and $N_{\rm H}$--$D$ relation, 
we were able to derive the values of $D$ and the J1952 emitting area radius $R$ independently, 
as opposed to their ratios obtained traditionally from such fits. 
For the NSMAX models, $R=10^{+21}_{-7}$~km ($B=10^{12}$~G) and $R=6^{+9}_{-3.5}$~km ($B=10^{13}$~G)
imply that the thermal emission  originates from the entire NS surface. 
For the BB model, the emitting area is significantly smaller, $R=0.6^{+0.6}_{-0.3}$~km, and the radiation 
likely comes from a hot polar cap of the pulsar.
To derive an upper limit on the  temperature of the entire NS surface in the latter case, 
we added a second BB component to the best-fit model. 
The radius of the emitting area was fixed at a conventional NS radius of 13~km. 
We obtained a 3$\sigma$ upper limit on the entire surface temperature of $\approx60$~eV.

For the  J1952 spectrum we also tried the PL model that provided
a worse fit quality ($\chi^{2}$/dof=489/461) and resulted in a too large photon index 
$\Gamma=4.4^{+0.6}_{-0.5}$, not typical for pulsars \citep*[see e.g.][]{Li2008}. 

Besides, we added a power law component to the thermal component for J1952 to estimate 
an upper limit on the  pulsar non-thermal luminosity $L^{\rm pl}_{\rm J1952}$. 
In current data this emission component is not resolved within the non-thermal PWN background. 
The photon index $\Gamma$ was assumed to lie in the range $0.5\leq \Gamma \leq2.0$ typical for other pulsars.
The resulting values of $L^{\rm pl}_{\rm J1952}$ are also presented in Table~\ref{t:best-fit}. 


\section{Discussion}
\label{sec:discus}

We used high temporal resolution \textit{XMM-Newton}/EPIC-pn archival data to search 
for pulsations from the NS J1952. 
No pulsations were detected.
However, we set a more stringent upper limit for the pulsed fraction of 40~per~cent 
(99~per~cent confidence) in a wider period range of $\ga 12$~ms (Sect.~\ref{sec:timing}) 
than previous estimates based on the \textit{ASCA} data \citep{arzoumanian2004}. 
We confirmed the result of \citet{arzoumanian2008} that only the thermal emission component is detected
in the X-ray spectrum of J1952 and better constrained its parameters.
For the BB model the emitting area radius is $\approx0.6$~km (see Table~\ref{t:best-fit}) showing 
that the thermal component originates from a small hot spot on the NS surface with $T\approx 215$ eV, 
which is typical for pulsar hot polar caps. 
On the other hand, the NSMAX models suggest that the emission comes from the entire NS surface.
We cannot assert definitely which case is true for J1952.
The detection of the pulsations, the phase-resolved spectral analysis and the analysis of 
pulse profile shapes could help to solve this problem.
Our upper limit on the PF of 40~per~cent is higher than typical PFs of 10--30~per~cent detected for
the thermal emission from entire surfaces of other NSs \citep{zavlin2007,pz2000}.

\begin{figure}
\centerline{
\includegraphics[scale=0.41]{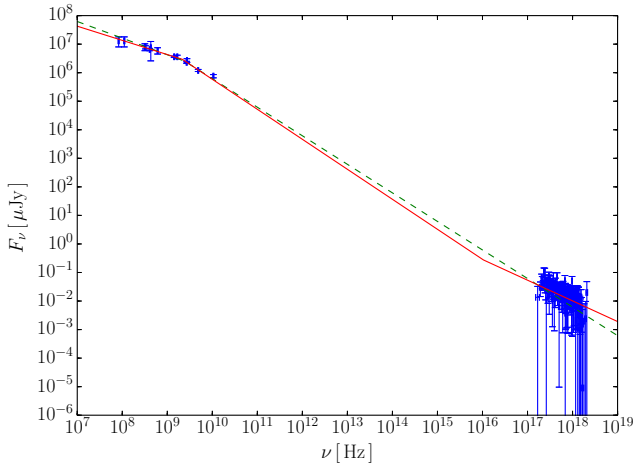}} 
\caption{Unabsorbed spectrum of the DA~495 PWN from the radio to X-rays. 
The X-ray spectrum of the associated pulsar J1957 is fitted by the BB model. 
The dashed and solid lines are the best fits to the PWN data by the broken PL model with the single 
and the double-knee breaks, respectively. The position of the second high frequency spectral break 
is poorly defined (see text for details), 
it is fixed here for illustrative purposes.} 
\label{fig:spec-pwn}
\end{figure}
 
The interstellar absorption--distance relation (Sect.~\ref{sec:relation}) 
allowed us to estimate the distance to DA~495. 
For the BB model $D=2.4^{+1.3}_{-1.1}$~kpc.
The NS atmosphere models suggest a larger distance: $3.3^{+1.7}_{-1.3}$~kpc, if the J1952 magnetic field 
is $10^{12}$~G, or $3.2^{+1.5}_{-1.2}$~kpc, if it is $10^{13}$~G (Table~\ref{t:best-fit}). 
Within uncertainties, all these values are consistent with one of the previous estimates, $D\approx 2.1$~kpc, 
assuming possible association of DA~495 with the open cluster NGC 6834 \citep{landecker1983}. 
Another distance estimate of $1.0\pm0.4$~kpc is compatible only with the BB model.  
It was based on the H{\sc\,i} absorption feature in the DA~495 polarized radio continuum emission with a systemic velocity 
of $+12$~km~s$^{-1}$, which is likely associated with the Galactic Local Arm \citep{kothes2008}. 
The authors also mentioned a distance of 5 kpc as an alternative interpretation of the feature. 
This is compatible at the upper confidence level with the result provided by our atmospheric fits. 
However, they argued that this distance is inconsistent with the low foreground
absorbing column density obtained by \citet{arzoumanian2008} from \textit{Chandra} X-ray data
using the BB model 
and with the absence of the H{\sc\,i} absorption from the Local Arm tangent point
expected to be at the velocity of about $+20$~km~s$^{-1}$. 
We note, that the latter absorption cannot be completely excluded due to a low signal-to-noise ratio
of the radio data. 
The atmospheric models cannot be also ignored in the above argument. 
Moreover, recent trigonometric parallax measurements of high mass star forming regions
show that the Local Arm extends beyond 5 kpc with a shallow pitch angle of $\approx10\degr$  
between $l\approx72\degr$ and $52\degr$ toward the Perseus Arm \citep{2014Reid, 2013Xu}. 
It is possibly a separate branch of the Perseus, connecting with it at a distance of about 6.5 kpc 
\citep[see fig. 12 from][]{2013Xu}. 
DA~495 with $l\approx65\fdg7$ is obviously located within the new Local Arm longitude range.  
Therefore, the gas within this arm can be a natural source of the larger foreground X-ray absorption 
at the larger DA~495 distance suggested by the atmospheric models (Table~\ref{t:best-fit}). 

The PWN spectral indices $\alpha_{\nu}$ derived from our X-ray fits are $0.71\pm0.12$,
$0.83\pm0.13$ and $0.82\pm0.13$ (90~per~cent confidence) when the thermal component from the NS 
was described by the BB and atmosphere models with $B=10^{12}$~G and $10^{13}$~G, respectively.
These values are in agreement within uncertainties with each other and with $\alpha_{\nu}=0.87\pm0.10$ 
(68~per~cent confidence) obtained from the radio data for frequencies above the 1.3 GHz break \citep{kothes2008}. 
However, a marginal spectral flattening in X-rays, particularly in the BB case, cannot be excluded, 
suggesting a second break. 
To examine whether an additional break between the radio and X-rays is needed or not,
we tried to fit the X-ray spectra together with the radio data taken from \citet{kothes2008}.  
At first, the PWN spectrum was fitted by a broken power law (BPL) with a single break.
In all cases the break in the radio was  $2.1^{+0.8}_{-0.5}$~GHz. 
This value is higher than that of \citet{kothes2008} due to the addition of the X-ray data; 
using only the radio data we reproduced their result of $\approx1.3$~GHz.  
The spectral indices $\alpha_{\nu1}$ and $\alpha_{\nu2}$ before and above this break were $0.61\pm0.07$ 
and $1.00\pm0.01$ for the NSMAX+BPL models and $0.58\pm0.07$ and $1.00\pm0.01$ for the BB+BPL model. 
Then we included a second break to the PWN model. 
For the NSMAX+BPL models, the resulting spectral indices $\alpha_{\nu1}$ and $\alpha_{\nu2}$ 
before and above the break at 2~GHz and the spectral index $\alpha_{\nu3}$ above the second break were
$0.50^{+0.09}_{-0.12}$, $1.02^{+0.07}_{-0.04}$ and $0.84^{+0.11}_{-0.12}$ in the case of $B=10^{12}$~G and 
$0.50^{+0.10}_{-0.13}$, $1.02^{+0.10}_{-0.04}$ and $0.83^{+0.15}_{-0.13}$ in the case of $B=10^{13}$~G.
For the BB+BPL model, $\alpha_{\nu1}=0.5^{+0.09}_{-0.11}$, $\alpha_{\nu2}=1.05^{+0.11}_{-0.05}$ and 
$\alpha_{\nu3}=0.72^{+0.11}_{-0.11}$.
We carried out an F-test using the {\sc xspec} {\sc ftest} routine to compare models
with single and double-knee breaks.  
The resulted probabilities for the NSMAX+BPL were $\approx 0.1$ for both values of magnetic field
demonstrating that the second break is not needed.
On the other hand, the probability for the BB+BPL model is $0.03$ providing a marginal evidence that the PWN 
spectrum in X-rays indeed becomes flatter although the second break position is poorly defined and lies in 
a range of $10^{14}-10^{17}$ Hz, somewhere between the mid-infrared and X-rays. 
The best fits to the data by the BB+BPL model with single 
and double-knee breaks are presented in Fig.~\ref{fig:spec-pwn}. 

Assuming that the spectral break at $\approx2$~GHz in DA~495 emission arises from the synchrotron cooling
and using the formula from \citet{ginzburg1965}, 
\begin{equation}
\label{eq:nu}
\nu_{\rm c}(\rm GHz) \approx 1.2\times 10^3 B_{\rm mG}^{-3}t^{-2}_{\rm kyr},
\end{equation}
where $\nu_{\rm c}$ is the cooling frequency, $B_{\rm mG}$ is the PWN magnetic field in milligauss, 
and $t_{\rm kyr}$ is the PWN age in kyr, we can estimate the age of DA~495.
The synchrotron lifetime is $\tau_{\rm synch}\approx 3.3R_{\rm pc}c/v$~yr, where $R_{\rm pc}$ is the 
emitting area radius in pc, $v$ is the velocity of the pulsar wind particles and $c$ is the speed of light. 
At a distance of about 1.5 kpc, which is close to the low boundary obtained from the X-ray spectral fits, 
$R_{\rm pc}=0.15D_{\rm 1.5 kpc}$, which corresponds to the 20~arcsec angular radius of the X-ray PWN,  
and $\tau_{\rm synch}\approx 1.7D_{\rm 1.5 kpc}$~yr, assuming a typical velocity $v=0.3c$ \citep{kennel1984}.
On the other hand, at an X-ray photon energy  $E=1.5$~keV,  
$\tau_{\rm synch}=1.2~E^{-1/2}_{\rm keV}B^{-3/2}_{\rm mG}\approx B^{-3/2}_{\rm mG}$.
This yields  $B_{\rm mG}\approx0.7D_{\rm 1.5 kpc}^{-2/3}$ and the expression~(\ref{eq:nu}) 
implies the DA~495 age $t_{\rm kyr}\approx 40D_{\rm 1.5 kpc}$. 
At the maximum distance of about 5~kpc we obtain $B_{\rm mG}\approx 0.3$, which is comparable 
to that of the Crab Nebula, and a maximum age $t_{\rm kyr}\approx 130$. 
These simplified estimates imply that DA~495 may be a factor of 2--6 older  
than it was suggested by \citet{kothes2008}. 

\begin{figure} 
\begin{minipage}[h]{1.\linewidth} 
\center{\includegraphics[scale=0.38]{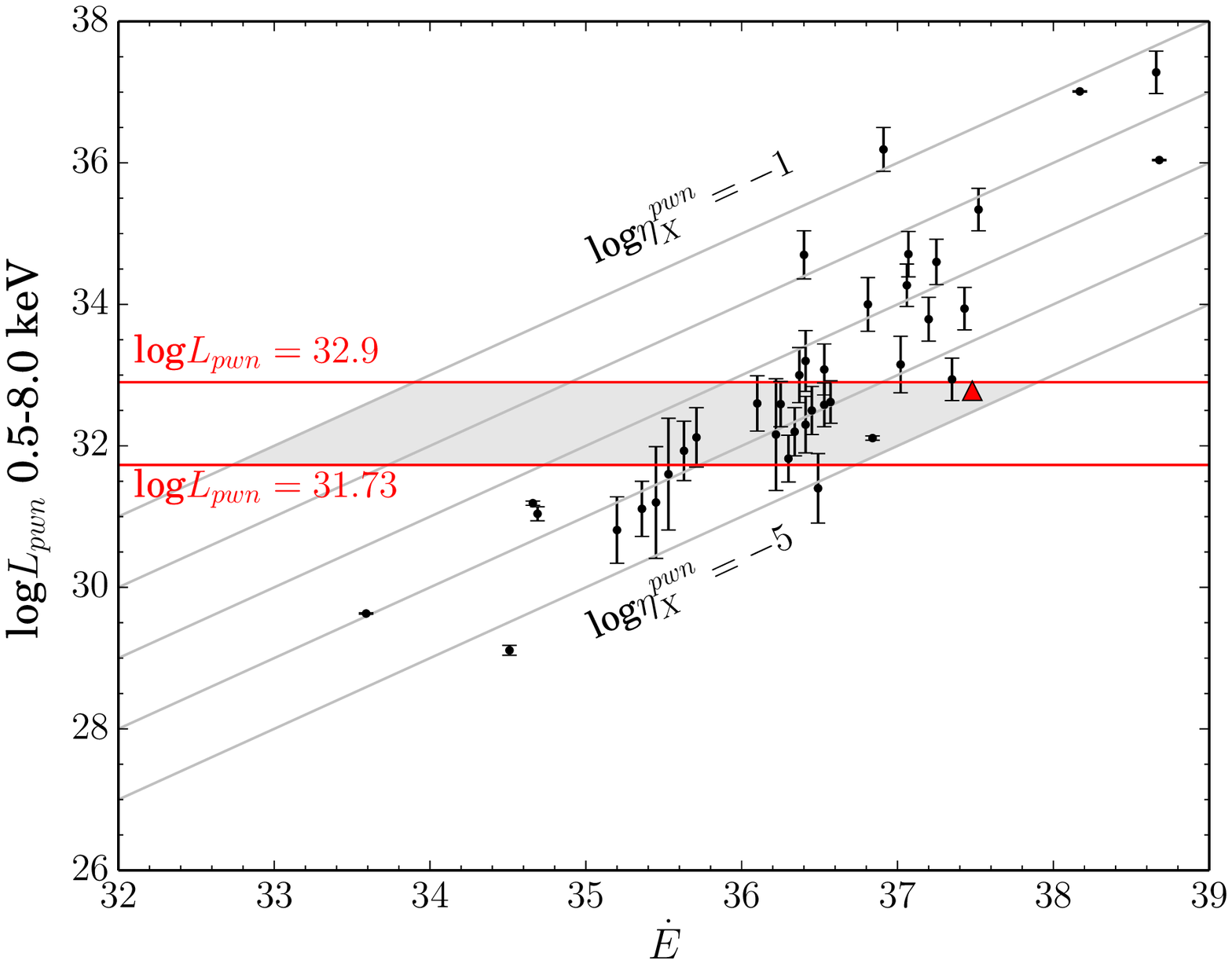}\\} 
\end{minipage} 
\begin{minipage}[h]{1.\linewidth} 
\center{\includegraphics[scale=0.38]{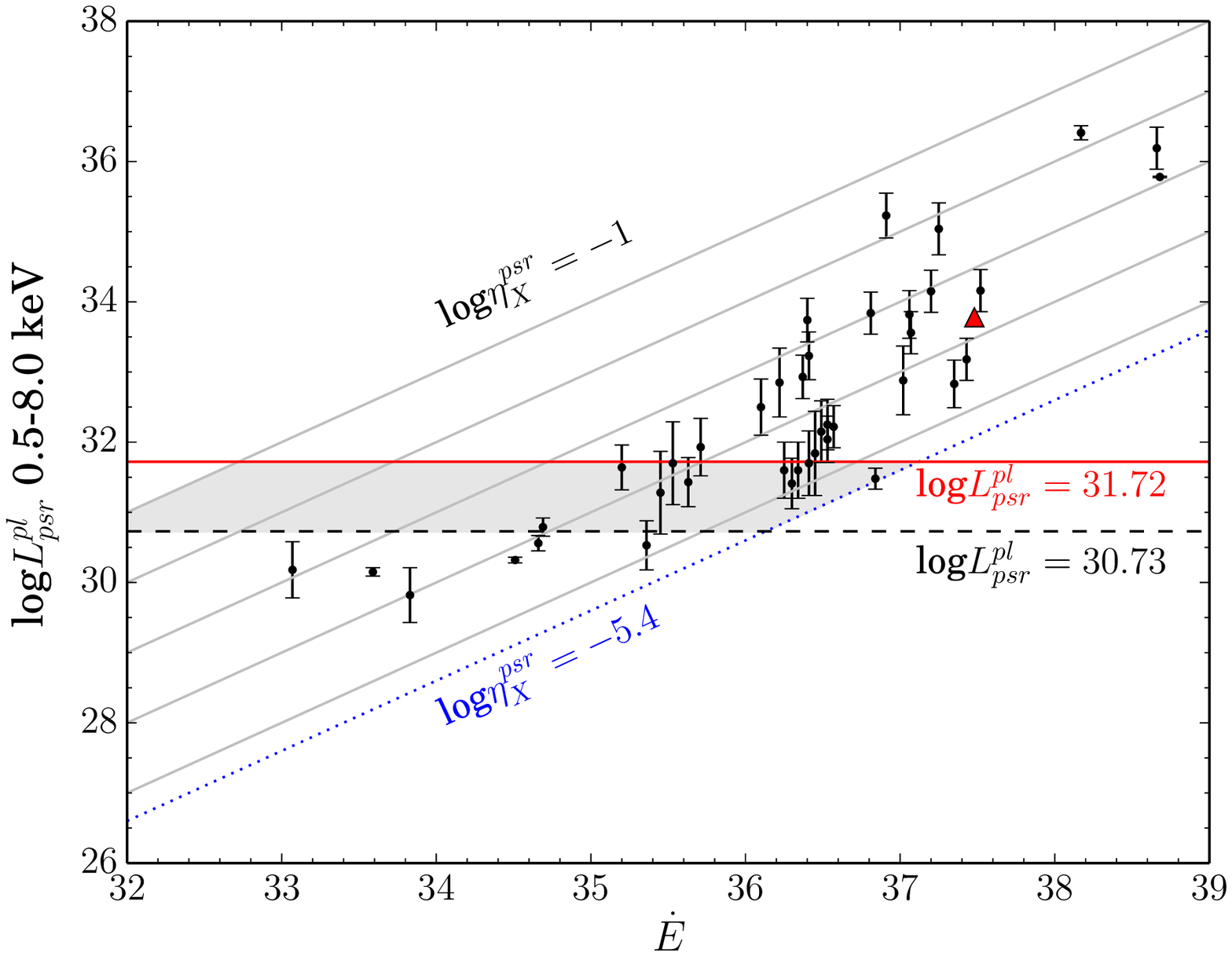}\\} 
\end{minipage} 
\begin{minipage}[h]{1.\linewidth} 
\center{\includegraphics[scale=0.38]{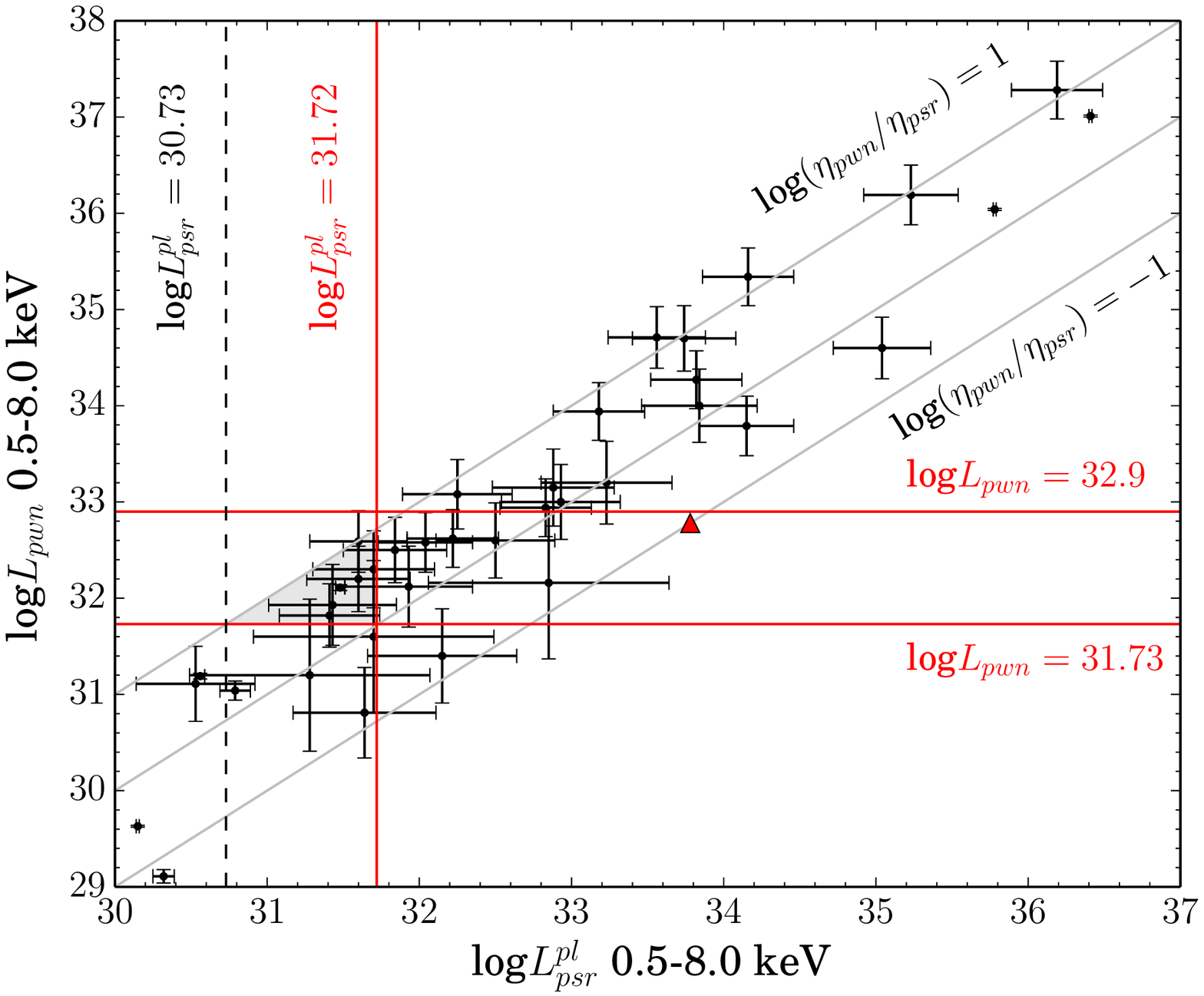}\\} 
\end{minipage}
\caption{ {\sl Top panel:}  X-ray luminosities of PWNe {\sl vs}  the spin-down luminosity $\dot{E}$. 
{\sl Middle panel:}  the same as is in the {\sl top panel} but for  X-ray non-thermal luminosities of 
the pulsars powering the PWNe.  
{\sl Bottom panel:} X-ray luminosities of the same PWNe {\sl vs} pulsar X-ray non-thermal luminosity. 
Error-bars include 40 per cent  systematic distance uncertainties for objects where parallax was not measured. 
Grey solid lines show constant efficiency levels $\eta$ for the PWNe and PSRs and their ratio 
as indicated in the plots. 
Horizontal solid lines correspond to the DA~495 PWN luminosity lower and upper bounds and the upper 
limit on the J1952 pulsar non-thermal luminosity obtained from the X-ray spectral fits. 
Their values are presented in the plots. 
The dotted line in the {\sl middle panel} indicates the lowest efficiency level corresponding to the Vela pulsar.
The black dashed lines represents the lower bound of the J1952 pulsar non-thermal luminosity.
The regions where the DA~495 PWN and its pulsar are expected to  locate are filled by the grey colour. }        
  
\label{fig:karg} 
\end{figure} 

\begin{figure*}
\begin{minipage}[h]{0.49\linewidth}
\center{\includegraphics[width=1.0\linewidth]{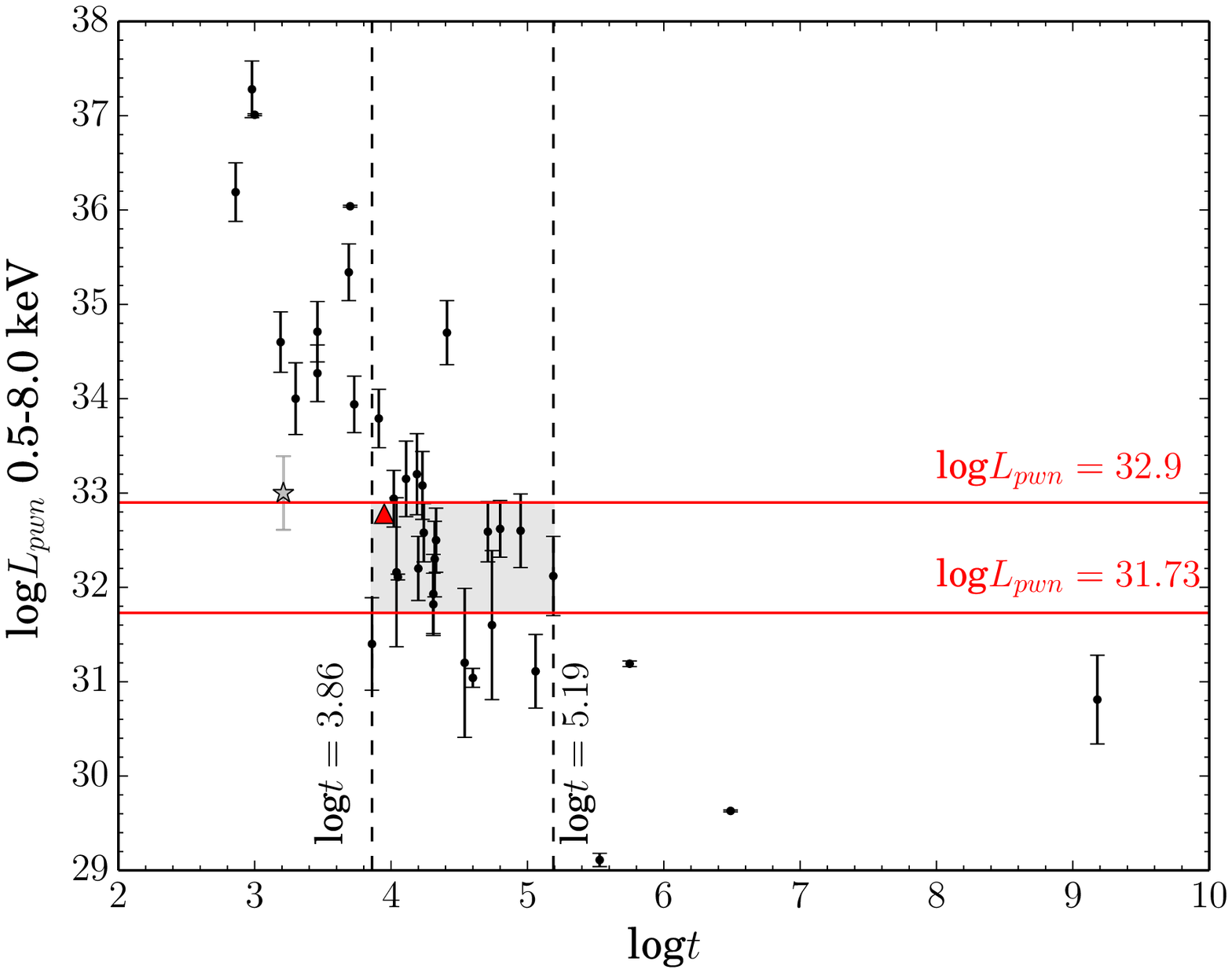}}
\end{minipage}
\hfill
\begin{minipage}[h]{0.49\linewidth}
\center{\includegraphics[width=1.0\linewidth]{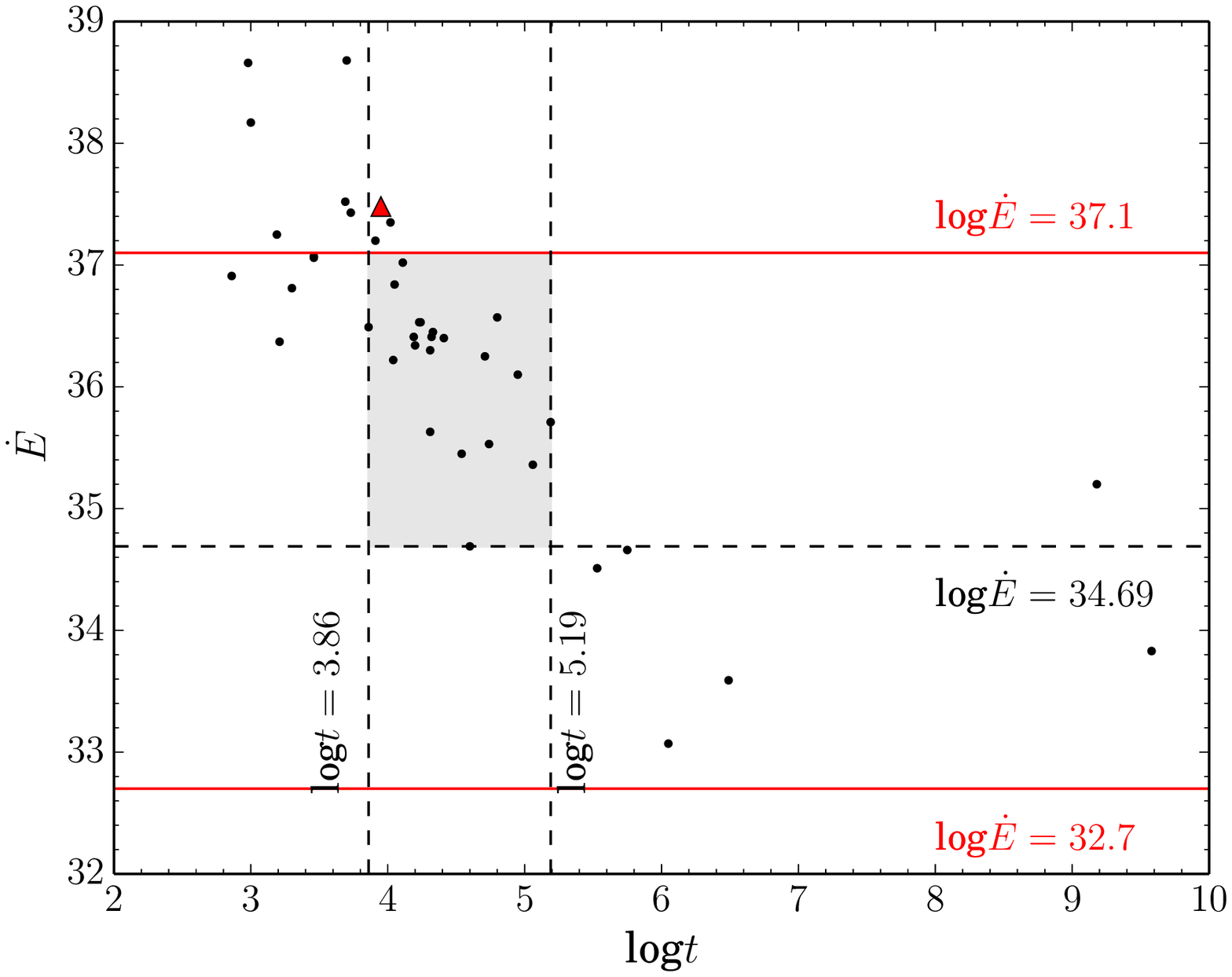}}
\end{minipage}
\caption{$L_{\rm pwn}$ for the same PWNe as in Fig.~\ref{fig:karg} ({\sl left}) and 
respective $\dot{E}$ ({\sl right}) {\sl vs} age t.
Solid lines show the DA~495 PWN luminosity 
bounds derived from the fits and its spin-down luminosity range estimated from 
$L_{\rm pwn}(\dot{E})$ and $L_{\rm psr}^{\rm pl}(\dot{E})$ dependencies presented in Fig.~\ref{fig:karg}. 
The vertical dashed lines show a possible age bounds for DA~495 and the horizontal dashed line in the {\sl right panel} demonstrates its possible spin-down luminosity lower bound based on the age bounds  and $\dot{E}(t)$ distribution for other pulsars. The regions where the DA~495 PWN and its pulsar are expected to locate are filled by the grey colour.}
\label{fig:karg2}
\end{figure*}

The distance estimate allowed us to derive the PWN luminosity $L_{\rm pwn}$ and the $3\sigma$
upper limit on the non-thermal luminosity of its pulsar $L_{\rm J1952}^{\rm pl}$ (Table~\ref{t:best-fit}). 
These parameters can be used to estimate the J1952 spin-down luminosity $\dot{E}$ 
utilizing the empirical dependencies $L_{\rm pwn}(\dot{E})$, $L_{\rm psr}^{\rm pl}(\dot{E})$ and 
$L_{\rm pwn}(L_{\rm psr}^{\rm pl})$ obtained for 0.5--8~keV band by \citet{kargaltsev2008} using the data 
for other pulsars+PWN systems (Fig.~\ref{fig:karg}). 
For completeness, we added to these data PSR J2022+3842 with its PWN G76.9+1.0
\citep{arzoumanian2011,arumugasamy2014} marked in Fig.~\ref{fig:karg} by a triangle. 
Summarising the results for all three models presented in Table~\ref{t:best-fit}, 
${\rm log}L_{\rm pwn}$ for DA 485 lies in the range of 31.7--32.9 (horizontal solid lines 
in the top and bottom panels of Fig.~\ref{fig:karg}), 
which implies ${\rm log}\dot{E}=$ 32.7--37.9 for the X-ray efficiency range 
$\eta_{\rm X}^{\rm pwn}=10^{-5}-10^{-1}$ based on the data for all PWNe observed 
(the grey shaded region between the horizontal solid lines in the top  panel). 
Using the respective empirical luminosity {\sl vs} age dependence $L_{\rm pwn}(t)$ for 
the same PWNe set (Fig.~\ref{fig:karg2}, left panel)\footnote{For age $t$, the pulsar spin-down 
ages are used except for the cases when the true ages are known from respective SNR studies: 
Crab, J1833$-$1034 \citep{Bietenholz2008}, J1811$-$1925 \citep{roberts2003}, 
B1951+32 \citep{migliazzo2002}, J0538$+$2817 \citep{ng2007}.}  and the DA~495 $L_{\rm pwn}$ allowed range, 
we can obtain the possible age range of DA~495 as $7 \la t_{\rm kyr} \la 155$ 
(the grey region between two vertical dashed lines in the left panel 
of Fig.~\ref{fig:karg2})\footnote{At this point  we excluded PSR J1119$-$6127, marked by the star symbol in 
the left panel of Fig.~\ref{fig:karg2}, as a peculiar high-B pulsar with a very faint PWN \citep{gonzalez2003}.}. 
This age range is slightly wider than that based on the simple synchrotron cooling scenario.  
At the same time, the derived upper limit on the pulsar nonthermal 
luminosity ${\rm log}L_{\rm J1952}^{\rm pl}<31.7$ 
(the horizontal solid line in the middle panel of Fig.~\ref{fig:karg}) 
suggests ${\rm log}\dot{E}<37.1$ for the pulsar efficiency range $\eta_{\rm X}^{\rm psr}\ga10^{-5.4}$ 
based on the data for all pulsars (the grey region in the middle panel of Fig.~\ref{fig:karg}).
The lower limit for $\eta_{\rm X}^{\rm psr}$ (the dotted line in the middle panel of Fig.~\ref{fig:karg}) 
corresponds to the Vela pulsar X-ray efficiency, which is known to be very under-luminous in X-rays.
Consequently, a real ${\rm log}\dot{E}$ value likely lies in the range of 32.7--37.1 where the allowed ranges of $\dot{E}$ for the PWN and 
J1952 overlap. 
This range boundaries are shown by solid 
lines in the $\dot{E}(t)$ empirical dependence in the right panel of Fig.~\ref{fig:karg2}.  
On the other hand, the $L_{\rm pwn}(L_{\rm psr}^{\rm pl})$ dependence allows us to set a lower limit of $\approx30.7$ on the J1952 
non-thermal luminosity (the vertical dashed line in the bottom panel of Fig.~\ref{fig:karg}).
This value is also shown by the horizontal dashed line in the middle panel of Fig.~\ref{fig:karg}.  
Finally, the $\dot{E}(t)$ empirical dependence in the right panel of Fig.~\ref{fig:karg2}  
and the estimated age range (the vertical dashed lines in this plot) suggest a more narrow 
${\rm log}\dot{E}$ range of 34.7--37.1 corresponding to the grey region in this panel between 
the upper solid and horizontal dashed lines. 
This appears to be the most reasonable  DA~495 $\dot{E}$ range summarising all data available for DA~495 
and other PWN+pulsar systems. 

Comparing thermal properties of J1952 derived from the X-ray spectral fits (Table~\ref{t:best-fit}) 
with NS cooling models \citep{yakovlev2004ARA}, we find that  the entire  surface temperature of 80--90 eV provided 
by the atmospheric models is  consistent with the standard cooling scenario for the obtained age range. 
For the BB model, a 3$\sigma$ upper limit on the entire surface temperature of $\approx$60~eV   
appears to be cooler than the standard cooling scenario predicts, if J1952 age is $\la 30$~kyr.

\citet{arzoumanian2011} have noticed, that by many properties DA~495 is similar to the PWN  G76.9+1.0 associated 
with the pulsar J2022+3842.
Both objects demonstrate bright radio PWNe with similar morphologies, while their X-ray PWNe are faint, 
have much smaller sizes than in the radio and coincide with central faint emission holes of the radio PWNe. 
The ratio of the PWN sizes in the radio and X-rays is about 20 for both objects.  
As for DA~495, no SNR shell was found for G76.9+1.0. 
PSR J2022+3842 with $P=48.6$ ms, detected in the radio and X-rays, has a characteristic age $\approx8.9$~kyr,  
$B\approx 2.1\times 10^{12}$ G and $\dot{E}\approx 3\times10^{37}$~erg~s$^{-1}$.  
Its PWN X-ray luminosity is $\approx 6\times10^{32}$~erg~s$^{-1}$, assuming the distance of 10 kpc     
\citep{arzoumanian2011,arumugasamy2014}.  
Positions of PSR J2022+3842 and its PWN in Fig.~\ref{fig:karg}, \ref{fig:karg2} are shown by triangles.
Given the similarity between DA~495 and G76.9+1.0, one can suggest that DA~495 may also be 
as young as G76.9+1.0, which is consistent with the low boundary of the age range predicted by the 
$\dot{E} - L_{\rm pwn}$ relation (see Fig.~\ref{fig:karg2}).
For the simple synchrotron cooling model the minimum age  
is about 30 kyr at the minimum allowed distance of 1.3 kpc inferred from the X-ray fits. 
This demonstrates that the simple synchrotron cooling model does not work in this case.
This is not a surprise, as it cannot explain spectra of other PWNe also and is used only for 
rough estimates \citep[e.g.][]{gaensler2006}.   

\begin{figure} 
\begin{minipage}[h]{1.\linewidth} 
\center{\includegraphics[scale=0.41]{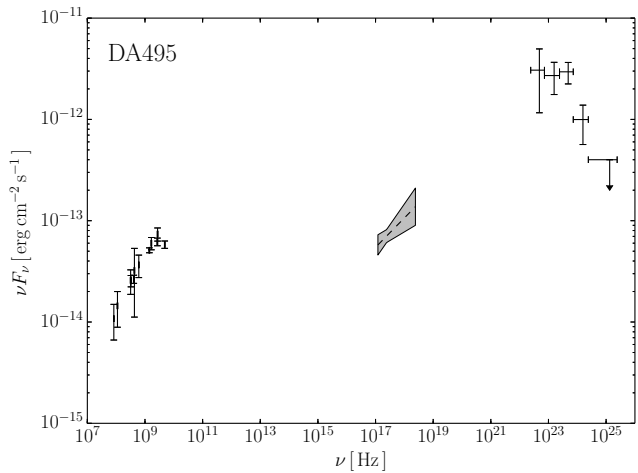}\\} 
\end{minipage} 
\caption{Radio and X-ray SED of the DA~495 PWN and the GeV spectrum of the associated \textit{Fermi} source 3FGL J1951.6+2926.}        
\label{fig:Fermi} 
\end{figure}

\textit{Fermi}/LAT recently detected at 6$\sigma$ significance a $\gamma$-ray source 3FGL J1951.6+2926  
coinciding spatially with DA 495 \citep{3Fermi}.  
Its spectrum is described by a power law with an exponential 
cut-off with an integral flux  $\approx10^{-11}$ erg cm$^{-2}$ s$^{-1}$ above 100 MeV.  
It was proposed as a possible DA 495 PWN $\gamma$-ray counterpart. 
However, comparing the source spectrum with the radio-X-ray spectrum  
of DA 495 compiled by us above,  
we find that such interpretation is hardly possible, since it would yield an unrealistically strong high energy excess 
in the PWN spectral energy distribution (SED) (Fig.~\ref{fig:Fermi}) which is not predicted by  any model of the PWN spectral evolution and not observed for most of evolved PWNe \citep[e.g.,][]{temim13,temim15}.  

An exception is a composite SNR MSH~11-62 containing the PWN where similar excess is observed \citep{Slane2012MSH1162}.
To explain the excess with  SNR models one needs to assume 
an unrealistically high ambient density for a hadronic scenario
or an unreasonably large fraction of the SNR mechanical energy converted to relativistic particles for a leptonic scenario.
The PWN model, in turn, requires an extremely high pulsar spin-down power. 
Based on that, \citet{Slane2012MSH1162} suggested that such a strong $\gamma$-ray emission of the SNR likely arises from its pulsar.  

\begin{figure} 
\begin{minipage}[h]{1.\linewidth} 
\center{\includegraphics[scale=0.41]{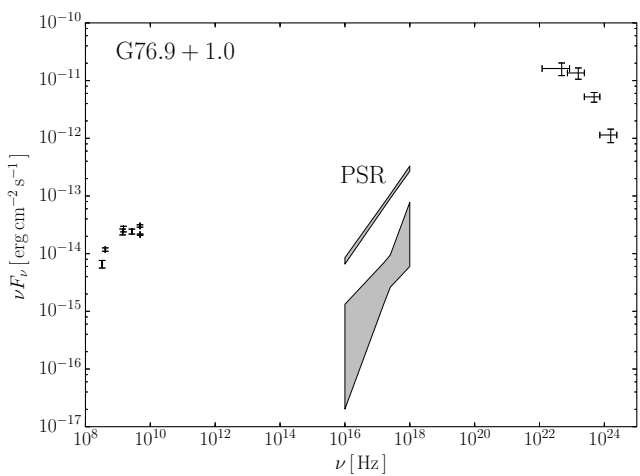}\\} 
\end{minipage} 
\caption{Radio and X-ray SED of the G76.9+1.0 PWN and the GeV spectrum of 3FGL J2022.2+3840, presumably the associated pulsar J2022+3842.
The  X-ray spectrum of the  pulsar,  marked by PSR, is shown for comparison. The  $\gamma$-ray emission level in respect to the radio-X-ray SED  
is similar to that of DA 495.
}        
\label{fig:G76} 
\end{figure}

Another unusual example is  G76.9+1.0, the twin of DA 495 discussed above. \citet{Martin2014} modelled its spectrum   
and concluded that the detection of this PWN at the GeV--TeV range seems unexpected.  Nevertheless, \citet{3Fermi} have reported on a possible 
$\gamma$-ray counterpart, 3FGL J2022.2+3840, 
of the G76.9+1.0 PWN. We find, that this implies similar unreasonably high energy excess 
in its SED as for  DA 495 and MSH~11-62 (Fig.~\ref{fig:G76}).   
At the same time, \citet{FermiJ2022} have reported on a $5\sigma$ detection of the pulsed  
emission from PSR J2022+3842, associated with this PWN, at energies above 200 MeV with  a pulsed flux  
of (2.7$\pm$0.5)$\times10^{-7}$ photons cm$^{-2}$ s$^{-1}$. 
This flux is compatible with that reported by  \citet{3Fermi}  for 3FGL J2022.2+3840, suggesting 
that both groups observed  the same object.  
For the distance of 10 kpc,  
this yields the pulsar luminosity $L_{\gamma}=(3.6\pm0.6) \times 10^{35}$ erg s$^{-1}$ and the $\gamma$-efficiency  
$L_{\gamma}/\dot{E}\approx 0.01$. 
3FGL J2022.2+3840 has also the power law spectrum with the exponential 
cut-off typical for other \textit{Fermi} pulsars \citep{abdo13}.


The DA 495 SED similarity with those of G76.9+1.0  and MSH~11-62 implies that the associated \textit{Fermi} source 3FGL J1951.6+2926   
is likely the $\gamma$-ray counterpart of the J1952 pulsar but not of the PWN. 
This is also supported by the pulsar-like spectrum of the source. 
In this case, the pulsar luminosity $L_{\gamma} \approx 10^{34}$ erg s$^{-1}$ for the distance of 3~kpc. 
Given that, and  assuming the same $\gamma$-efficiency as for PSR J2022+3842, 
we obtain $\dot{E}\approx 10^{36}$~erg~s$^{-1}$ for J1952, which is within the range 
obtained above using the X-ray data.  

\section{Summary}
\label{sec:sum}

The X-ray spectrum of J1952 is pure thermal and can be  described  either by the BB or NS atmosphere models. 
In the former case the emission comes from a pulsar hot spot and in the latter -- from a bulk of the NS surface 
with $T \approx$85 eV consistent with standard NS cooling scenarios.  
The $N_{\rm H}$--$D$ relation provides independent constraints on the distance to the object.
For atmosphere models it can be larger, up to 5 kpc, than it was supposed early. 
The upper limit for the pulsed emission fraction of J1952 is 40~per~cent at 99~per~cent confidence.
It does not help to discriminate whether the thermal emission originates from the hot spot or the bulk of 
the NS surface.
The derived PWN X-ray luminosity range and the pulsar nonthermal luminosity upper limit 
combined with those of other pulsar+PWN systems imply the DA~495 age of 7--155 kyr. 
Empirical relations between $\dot{E}$, $L_{\rm pwn}$, $L_{\rm psr}^{\rm pl}$ and age
predict a possible J1952 ${\rm log}\dot{E}$ range of 34.7--37.1.
The BB model for J1952 suggests that the PWN spectrum likely becomes flatter in X-rays than in 
the radio above a break near 2 GHz, suggesting a second high frequency spectral break. 
The atmosphere models do not require the second break. 
Sub-millimetre and  infrared observations could help to check that. 
We suggest, that 3FGL J1951.6+2926 is the likely $\gamma$-ray counterpart of the J1952 pulsar. 
To constrain J1952 energetics and  age, deeper X- and $\gamma$-ray timing observations are needed.


\section*{Acknowledgments}

The scientific results reported in this article are based on data obtained from the 
\textit{Chandra} Data Archive and observations obtained with \textit{XMM-Newton}, an ESA science mission 
with instruments and contributions directly funded by ESA Member States and the USA (NASA). 
We thank the anonymous referee for useful comments and Dmitrii Barsukov and Denis Baiko for helpful discussion.
The work of AK and DZ was supported by RF Presidential Programme MK-2837.2014.2.
YS acknowledges the support from Russian Foundation for Basic Researches (grant 14-02-00868-a).

{\it Facilities: CXO, XMM}. 

\bibliographystyle{mnras}
\bibliography{refDA495}

\label{lastpage}

\end{document}